# Derivation of UML Based Performance Models for Design Assessment in a Reuse Based Software Development Approach

**Jasmine K.S[1], Dr. R. Vasantha [2]**
[1] Dept. of MCA, R.V.C.E., Bangalore-560 059, India
jasminesadeep@yahoo.co.in
[2] Dept. of information science & Engineering, R.V.C.E., Bangalore-560 059, India
vasanthaprak@yahoo.com

**ABSTRACT**. Reuse-based software development provides an opportunity for better quality and increased productivity in the software products. One of the most critical aspects of the quality of a software system is its performance. The systematic application of software performance engineering techniques throughout the development process can help to identify design alternatives that preserve desirable qualities such as extensibility and reusability while meeting performance objectives. In the present scenario, most of the performance failures are due to a lack of consideration of performance issues early in the development process, especially in the design phase. These performance failures results in damaged customer relations, lost productivity for users, cost overruns due to tuning or redesign, and missed market windows. In this paper, we propose UML based Performance Models for design assessment in a reuse based software development scenario.

## Introduction

The design and construction of future software systems will require the integration of software analysis and design methods with Software

163

Performance Engineering (SPE) in reuse based software development. This integration allows software designers to explore design alternatives and select a design that provides the best overall combination of understandability, reusability, modifiability and performance so that software systems can meet performance goals [Man98]. Central to improve the practice of performance implementation is the understanding that good design and management of resources will avoid the component communication bottleneck and performance failures.

Effective planning and analysis in the early stage of development process enables the organization to identify what type of practices is required for their products and plan ahead of time [V+05].

## 1. Reuse-based software development

The reuse approach to software development has been used for many years. However, the recent emergence of new technologies has significantly increased the possibilities of building systems and applications from reusable components. Large scale component reuse leads to savings in development resources, enabling these resources to be applied to areas such as quality improvement. Experience has shown that reuse-based development requires a systematic approach to and focus on the component aspects of software development [HRS95]. There are a number of software engineering disciplines and processes, which require specific methodologies for application in reuse-based development.

Current thinking is that the progress of software development in the near future will depend very much on the successful establishment of reuse-based development and this is recognized by both industry and academia.

### Software Performance Engineering (SPE)

SPE is a method for constructing systems to meet performance objectives [LFG05]. The process begins early in development and uses quantitative techniques to identify satisfactory designs and to eliminate those that are likely to have unacceptable performance before developers invest significant time in their implementation. SPE continues through the detailed-design, implementation and testing phases to predict and manage the performance of the evolving software and to monitor and report actual performance against specifications and predictions.

In particular, performance properties are essential in the context of reuse oriented and component based software production for two basic reasons [B+02]:

1. Among the multiple component implementations providing the same functional behavior, the clients will choose those components that best fit their performance requirements.

2. If components have performance specifications, then the performance of the system can be compositionally derived based on its components, while the component implementations need not be re-analyzed in each new context where they are used.

Our research work aims at developing a design based, implementation independent performance guaranteed software product by combining the most recent advances in the fields of: (i) Reuse based software development (ii) Software Performance Engineering (SPE) and (iii) UML modeling of software systems design. Our basic idea is to adapt the SPE approach to reuse based software development in the design phase to guarantee specific performance requirements.

**Present state in software reuse world & SPE**

In the research community, there are notable approaches to software performance engineering. Recent interest in software architectures has underscored the importance of architecture in achieving software quality objectives, including performance [Man98] [Smi90]. While decisions made at every phase of the development process are important, architectural decisions have the greatest impact on quality attributes such as modifiability, reusability, reliability, and performance [S+01] [Lav83].

The methodology for performance engineering demands extra effort and capabilities. Much recent researches are aimed at automating the performance modeling process [HW91] [WW04] [CM02]. But there is a need to specify performance parameters in these models. It requires skilled people. Our research aimed at facilitating this modeling process in the design level with the help of most widely used software-modeling language, namely unified modeling language (UML). Consequently UML diagrams, especially sequence diagrams, collaboration diagrams, activity diagrams, state machine (chart) diagrams and deployment diagrams play an important role in this process.

## 2. Implementation

### 2.1. SPE Models for performance Prediction

Software performance engineering is a quantitative approach to constructing software systems that meet performance objectives. It incorporates models for representing and predicting performance as well as a set of analysis methods, techniques for gathering data, and other steps mentioned earlier. Work on the creation of performance models from design notations includes [LFG05], [LJE98], [B+02].

Two types of models can be used to provide information for design assessment: the software execution model and the system execution model. The software execution model represents key aspects of the software execution behavior. This is a static model, which gives static performance measures. It is constructed using an execution graph to represent each performance scenario [SW01]. Nodes represent components of the software; arcs represent control flow. The graphs are hierarchical with the lowest level containing complete information on estimated resource requirements. If the software execution model indicates that there are no problems, analysts proceed to construct and solve the system execution model. The system execution model represents the key computer resources as a network of queues. This is dynamic model, which gives the dynamic performance measures. Queues represent components of the environment that provide some processing service, such as processors or network elements. Environment specifications provide device parameters (such as CPU size and processing speed). Workload parameters and service requests for the proposed software come from the resource requirements computed by solving the software execution model.

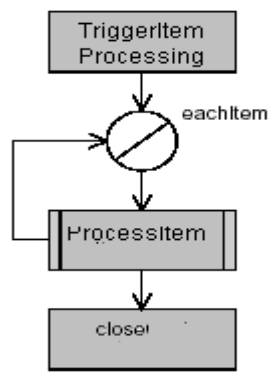

Fig.1 A Sample execution graph

In the figure 1, the graph shows that, following the node, TriggerItemProcessing, the ProcessItem node will repeat until each item completes it's processing. If the numbers of repetitions are known, we can label with the number of times instead of the label each item. From the difference in notation, it is clear that the node, ProcessItem can be expanded as a separate graph.

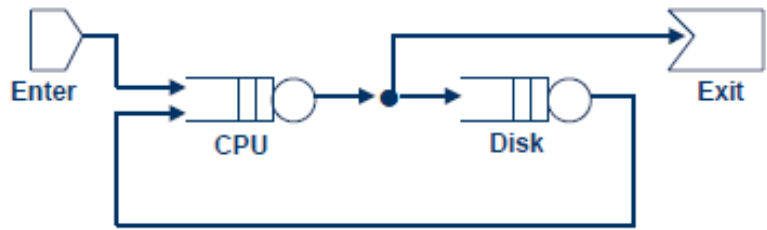

Fig.2 A Simple Queueing Network model (QNM)

There are 2 types of QNM: Open and Closed models:
Open: jobs enter & leave.
Closed: no arrivals/departures.

## 2.2. Discussion of UML diagrams in the Construction of Performance models

The SPE process begins with the system's use cases [Kaz et.al96] . Use cases describe the major functionalities of the system. Here we focus on the scenarios that describe the use cases. The scenario shows the objects that participate and the messages that flow between them. Performance scenarios are the subset of the use case scenarios that are executed frequently, or those that are critical to the perceived performance of the system. We use Unified Modeling Language (UML) sequence diagrams (SD) to represent performance scenarios. The SD objects represent the components involved, and the SD messages represent the requests of execution of a component service or correspond to information/data exchanged between the components.

Figure 3 depicts a sample sequence diagram. We can show synchronous and asynchronous messages in the UML using different types of arrowheads. In Figure 3 the communication between CompB and CompC is a synchronous communication and between CompC and CompD is an asynchronous communication. Also CompD has a self-delegation loop. All

167

these examples use standard UML notations. Additional extensions to the sequence diagram notation are in [UCD05].

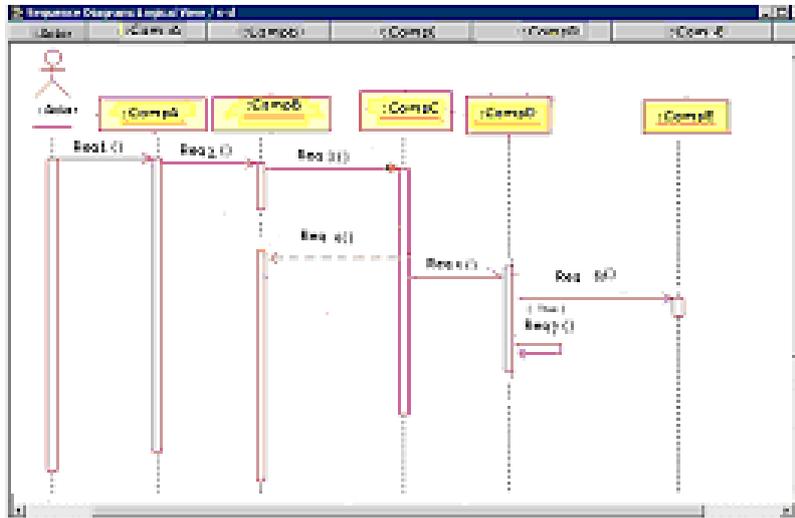

Fig3: Sequence Diagram

## 2.3 Deriving Execution Graphs from Sequence Diagrams

The performance analysis techniques are based on execution graphs. Thus, a key step in the SPE process is the derivation of execution graphs from sequence diagrams [SW01]. For scenarios with sequential flow of control, going from a sequence diagram to an execution graph is straightforward. For scenarios that occur parallel, a little more effort is needed to identify operations that serialize and account for communication and synchronization delays. In either case, the process of translating a sequence diagram to an execution diagram is similar.

Each message received by an object triggers an action - either an operation or a state machine transition. The simplest way to construct an execution graph from a sequence diagram is to follow the message arrows through the performance scenario and make each action a basic node in the execution graph. However, in many cases, individual actions are not interesting from a performance perspective and several of them may be combined into a single basic node. Alternatively, you can use an expanded node to summarize a series of actions and provide details of the sequence of actions in its sub graph. The walk, through the scenario will identify repetitions.

**2.4 Deriving Execution Graphs from Activity Diagrams**

A UML activity diagram shows the operational workflow of a system i.e., it will tell us which activities are executing sequentially and concurrently.

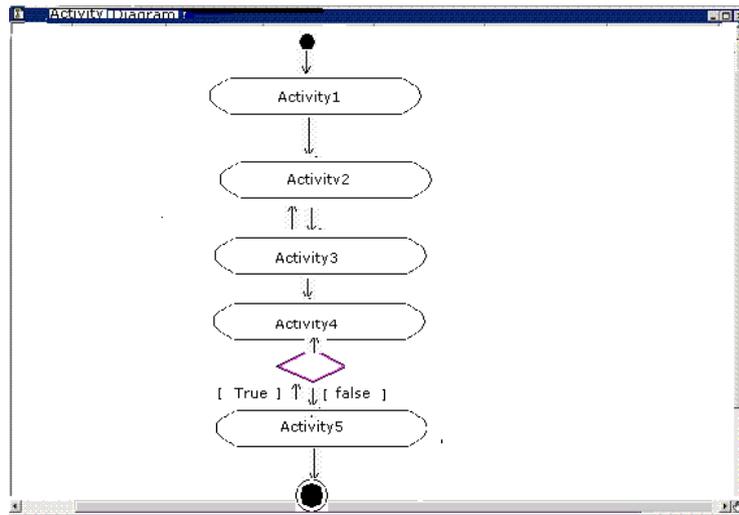

Fig4: Simple Activity Diagram

In fig 4, activity 1 to activity 4 is sequential in nature. Then a condition check is taking place, if the condition is true (corresponding to the self-delegation loop in sequence diagram, control will go back to action 4 itself. If the condition is false, the control will go to activity5.

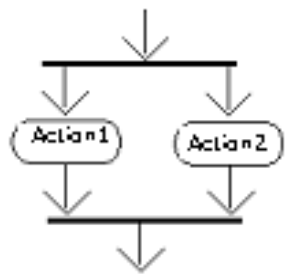

Fig.5. Activity diagram depicts concurrent activities

169

In fig 5, action 1 and action 2 are concurrent activities. For activities, which occur sequentially, going from an activity diagram to an execution graph is straightforward. For activities, which occurs concurrently, a little more effort is required.

**2.5 Deriving Execution Graphs from Collaboration Diagram**

UML collaboration diagram describes how the software components interact. An illustration is given in fig 6.The transformation from a sequence diagram into a collaboration diagram is a bi-directional function. The difference between sequence diagrams and collaboration diagrams is that collaboration diagrams emphasize more on the structure than the sequence of interactions. Within sequence diagrams the order of interactions is established by vertical positioning whereas in collaboration diagrams the sequence is given by numbering the interactions. So the execution model derived with the help of sequences diagrams and activity diagrams can be refined with the help of collaboration diagrams. By observing the number of arrows leading to a particular component, the utilization of that component can be predicted. So the requests sent to that component by other components have to wait, therefore response time will be more for them, resulting in performance degradation.

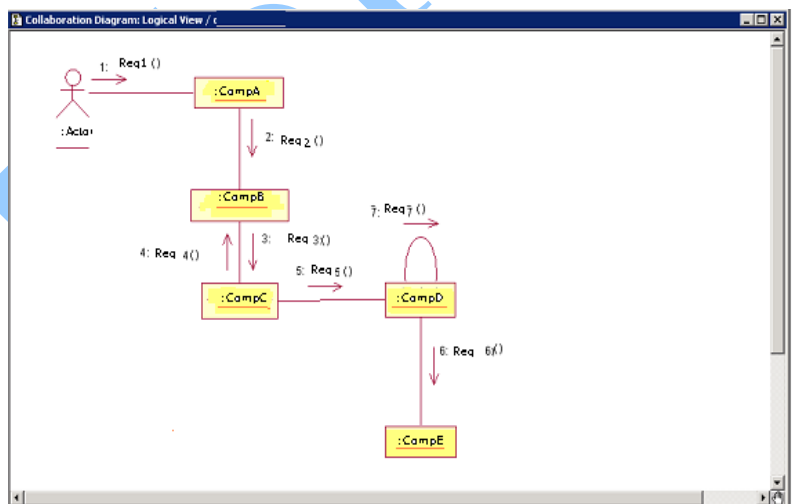

Fig.6.Collaboration diagram

In fig. 6, there is a two-way communication taking place between CompB and CompC.Also the CompD has to respond to CompC and it also

170

has a self-loop. So from the diagram, CompC and CompD are the most utilized component nodes compared to other component nodes. So the performance attributes of these components have to be monitored seriously.

### 2.6 Deriving Execution Graphs from State Chart Diagram

Solving the software execution model provides a static analysis of the mean, best and worst-case response times. If But we require a dynamic model that characterizes the software performance in the presence of factors, such as other workloads or multiple users that could cause contention for resources. So the system execution model has to be derived. The results obtained by solving the software execution model provide input parameters for the system execution model. State chart diagrams and deployment diagrams can play important roles in this direction. Solving the system execution model, the information about bottleneck resources and comparative data on options for improving performance through performance scenario changes, software changes and hardware upgrades are available.

State diagrams presents states of an object .It presents these state changes along with i)¨The transitions between the states ii) start point and end point of a sequence of state changes. Objects change the state in response to event or time. Each state represents the cumulative history of its behavior. Changes within the working state can be represented using sub states. In the case of sequential substates, translating state chart diagram to execution graph will be simple. But in the case of concurrent sub states, i.e., separating the working state into two components, a little more effort has to be taken.

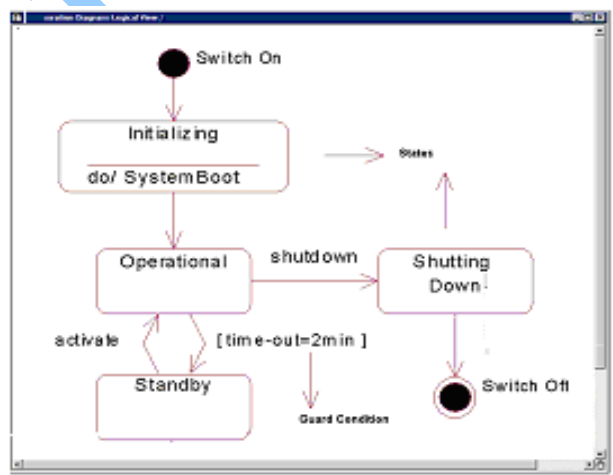

Fig.7. A sample state chart diagram

171

In fig 7, all states are sequential.

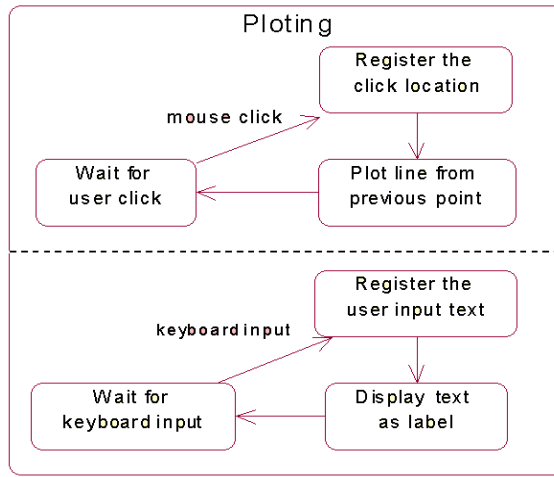

Fig. 8.State chart diagram depicts concurrent sub states

## 2.7 Deriving Execution Graphs from Deployment Diagram

A UML deployment diagram (DD) shows the allocation of the software components of the system to the processing nodes and the interconnection between the processing nodes (processes, workstations, I/O devices). The same diagrams can be re-used for similar applications, by only updating the associated parameters. The SD and DD diagrams have to be annotated with the proper performance values and parameters (PAs). For example, system and component execution times, response times, resource utilization (CPU utilization, disk, memory, network) I/O rates and average service time, network utilization, message size etc. A sample DD is shown in fig 8.

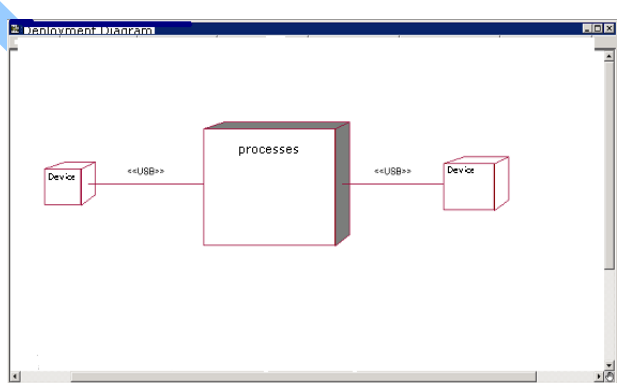

Fig.9.Deployment diagram

Considering the DD nodes, the PA attributes concern the resource scheduling policy (i.e. the strategy by which the resource handles the different jobs), the resource utilization and the resource throughput that represents the amount of work provided per unit of time by a resource belonging to a certain node. Also concurrency mechanism is needed for multiple instances of components. It can be implemented by configuration modification in the external deployment descriptors. So the deployment diagram can contribute in a large scale to build and solve system execution model.

**2.8 Model Overview**

Following are the steps in the construction and solving performance Models proposed in our study.

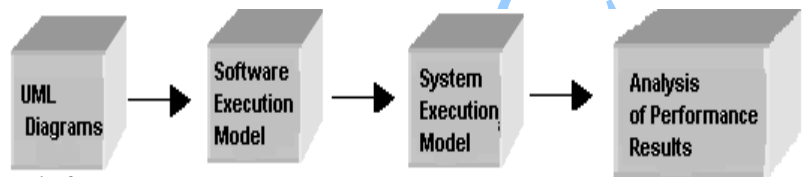

Fig.10.Block diagram of the Proposed Model

Input: Design specifications for the software product
Step1: Determine use cases and performance scenarios
Step2: Draw Sequence diagrams (SD) to identify parallel and sequential communication
Step3: Draw Activity diagrams to identify sequential and concurrent operational Workflow
Step 4: Generate Collaboration diagrams from SD to know the bottleneck components
Step5: Construct the software execution model and evaluate the statistical estimate of performance parameters
Step6: Draw state chart diagram to identify concurrent and sequential state machine transition
Step7: Draw Deployment diagram (DD) to know the allocation of software components of the system to processing nodes and their interconnection
Step8: Construct system execution model and evaluate the dynamic estimate of performance parameters

Step9: Analyze the performance models

Step10. review the proposed design and construct design alternatives in terms of the performance requirements based on the results provided by the performance models

Step11: Repeat the process throughout the development process

Output: Software product, which guarantee the required performance

Solving the software and the system execution models, the information about system and component execution times, response times, resource utilization (CPU utilization, disk, memory, network) thereby bottleneck resources, I/O rates and average service time, network utilization, and comparative data on options for improving performance through performance scenario changes, software changes and hardware upgrades etc will be available.

If the model results indicate that the performance is likely to be satisfactory, developers proceed. If not, the model results provide a quantitative basis for reviewing the proposed design and evaluating alternatives [SW93].

An ATM example:

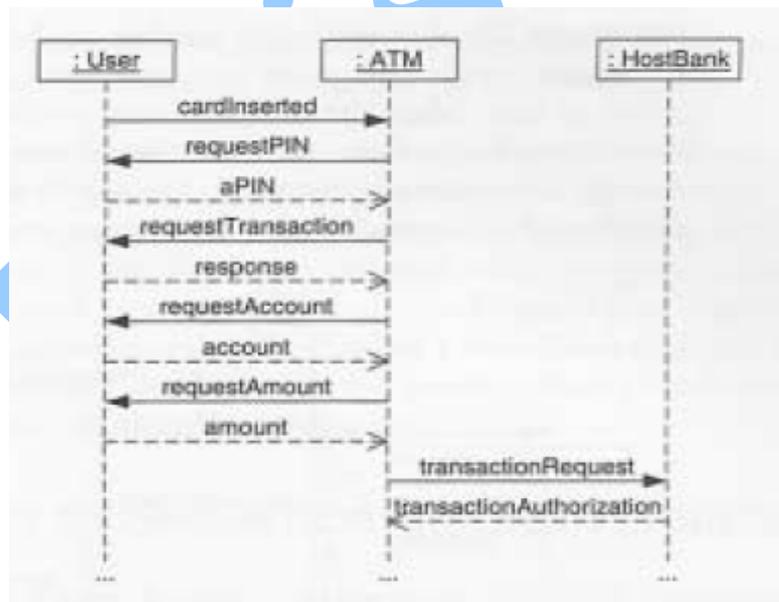

Fig.11. Sequence Diagram depicting normal transaction in an ATM machine

174

For the Fig. 13, by assuming n = 2, the shortest path, the longest path, and the average path for "Process Transaction" can be computed as follows:

Table: 1 Node Times for ATM calculation

| Node | Time Units |
|---|---|
| getCardinfo | 50 |
| getPIN | 20 |
| ProcessTransaction | 30 |
| processDeposit | 500 |
| processWithdrawal | 200 |
| processBalanceInquiry | 50 |
| terminateSession | 100 |

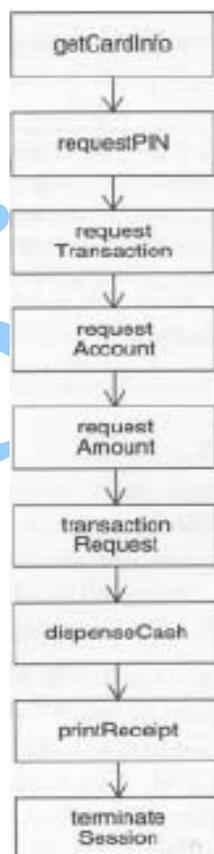

Fig.12.Corresponding Execution graph for fig.11

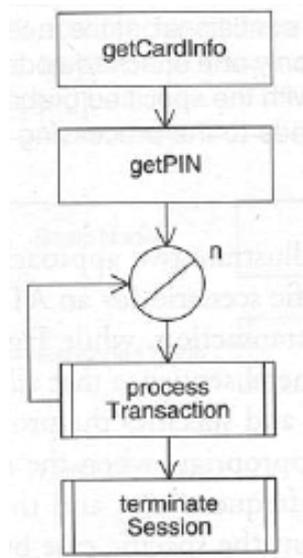

Fig.13. Execution graph for General ATM scenario

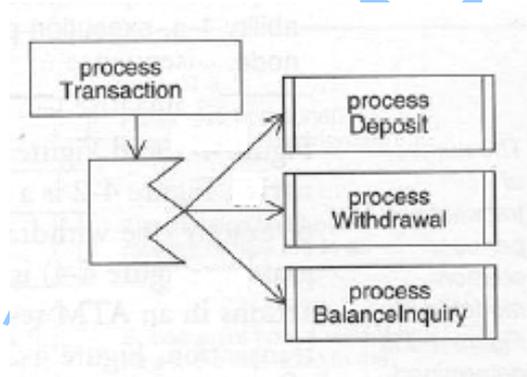

Fig14. Subgraph for process Transaction expanded node

In the table 1, time units are generated as random digits.
Shortest_path_PT = 80 time units
Longest_path_PT = 530 time units
Average_path_PT = 185.45 time units

Table: 2 Computer Resource Requirements for send Results

| Software Resource Requests | | Processing Overhead | | |
|---|---|---|---|---|
| Name | Service Units | CPU | Physical I/O | Network Messages |
| WorkUnit | 2 | 20 | 0 | 0 |
| DataBase | 1 | 100 | 2 | 0 |
| Messages | 1 | 5 | 2 | 1 |
| Send Results | | 400 | 3 | 1 |

Table 2 describes another System Execution model for computer resource requirements for each software resource request.

**Conclusion**

Models based on data flow are adequate for a view of a system in execution. It is not feasible to design complex systems from this basis. Representations which characterize units by the services they provide and specify behavior in a manner which is abstract but unambiguous are required. UML models and execution models which would enable a class of system tradeoffs which require means to assess performance and robustness play an important role in this direction. Our research work relies on, the most recent advances in the fields of: (i) Reuse based software development (ii) Software Performance Engineering (SPE) and (iii) UML modeling of software systems design. Our basic idea is to adapt the SPE approach to reuse based software development in the design phase to guarantee specific performance requirements. Further, refining the design and constructing more detailed models or constructing performance benchmarks and measuring resource requirements for key components can provide more precision to the predicted performance estimates. Future work can propose an automated environment for implementation of the steps mentioned and its application to case studies coming from the industrial world.

**Acknowledgment**

The work described in this paper was with the support and collaboration of ideas of several industry people residing in and around Bangalore. We wish to thank them for their contribution towards the successful completion of this research.